\documentclass[twocolumn]{aastex631}

\usepackage{amsmath}
\usepackage{amssymb}
\usepackage{natbib}
\usepackage{enumitem}
\usepackage{nicefrac}

\usepackage{savesym}
\savesymbol{tablenum}
\usepackage{siunitx}
\newcommand{\diff}[1]{\operatorname{d}\!{#1}}
\newcommand{\deriv}[3][0]{\frac{\operatorname{d}\if#10\else^{#1}\!\fi\!{#2}}{\operatorname{d}\!{#3}\if#10\else^{#1}\fi}}
\newcommand{\sderiv}[3][0]{\operatorname{d}\if#10\else^{#1}\fi\!{#2}/\operatorname{d}\!{#3}}
\newcommand{\pderiv}[3][0]{\frac{\partial\if#10\else^{#1}\!\fi{#2}}{\partial{#3}}}
\newcommand{\pcderiv}[4][0]{\left(\frac{\partial\if#10\else^{#1}\!\fi{#2}}{\partial{#3}}\right)_{#4}}
\newcommand{\spderiv}[3][0]{\partial\if#10\else^{#1}\fi{#2}/\partial{#3}}

\newcommand{\npg}{n_{\rm pg}}

\newcommand{\vy}{\mathbf{y}}

\newcommand{\wC}{\mathcal{C}}
\newcommand{\wN}{\mathcal{N}}
\newcommand{\wG}{\mathcal{G}}
\newcommand{\wT}{\mathcal{T}}

\newcommand{\relerr}{\Delta}

\newcommand{\gyre}{GYRE}

\newcommand{\BV}{Brunt-V\"ais\"al\"a}

\graphicspath{{./}{figures/}}

\begin{document}

\title{Corrected Weight Functions for Stellar Oscillation Eigenfrequencies}

\author[0000-0002-2522-8605]{R. H. D. Townsend}
\affiliation{Department of Astronomy, University of Wisconsin-Madison, 475 N Charter St, Madison, WI 53706, USA}
\email{townsend@astro.wisc.edu}
\author[0000-0002-6536-6367]{S. D. Kawaler}
\affiliation{Department of Physics and Astronomy, Iowa State University, 2323 Osborn Drive, Ames, IA 50011, USA}
\email{sdk@iastate.edu}

\begin{abstract}
\citet{Kawaler:1985} present a variational expression for the
eigenfrequencies associated with stellar oscillations. We highlight
and correct a typographical error in the weight functions appearing in
these expressions, and validate the correction numerically.
\end{abstract}


\section{Introduction} \label{s:intro}

Under the assumption of a vanishing surface density, the equations
governing linear, adiabatic, non-radial stellar oscillations are
self-adjoint and therefore obey a variational principle \citep[see,
  e.g.][]{Unno:1989}. Variational expressions for mode
eigenfrequencies serve as a valuable computational check on numerical
solutions to the oscillation equations. Via so-called weight
functions, they also quantify how eigenfrequencies are sensitive to
different parts of a star's internal structure; they are therefore key
tools in determining the asteroseismic probing potential of each mode.

\citet{Kawaler:1985} present a variational expression that has been
adopted in many studies of stellar oscillations
\citep[e.g.,][]{Kawaler:1986,Brassard:1992a,Brassard:1992b,Corsico:2002,Timmes:2018}. In
the present paper we highlight and correct a typographical error in
one of the weight functions appearing in their expression, and use
numerical calculations to demonstrate the validity of the correction.

\section{Weight Functions}

The \citet{Kawaler:1985} variational expression for the eigenfrequency
$\sigma(\vy)$ of a mode is
\begin{equation} \label{e:var}
  \sigma^{2}(\vy) = \frac{
    \int_{0}^{R} \left[
      \wC(\vy,r) + \wN(\vy,r) + \wG(\vy,r)
      \right]
    \rho r^{2} \diff{r}
  }{
    \int_{0}^{R}
      \wT(\vy, r)
      \rho r^{2} \diff{r}
  }.
\end{equation}
Here, $\vy(r) = \{y_{1}(r),\ldots,y_{4}(r)\}$ is the vector of
eigenfunctions introduced by \citet{Dziembowski:1971}, $r$ is the
radial coordinate (with $R$ the stellar radius), and $\rho$ the
density. The three weight functions are
\begin{align}
  \label{e:wC}
  \wC(\vy,r) &= g^{2} \ell(\ell+1) S_{\ell}^{-2} (y_{2} - y_{3})^{2}, \\
  \label{e:wN}
  \wN(\vy,r) &= r^{2} N^{2} y_{1}^{2}, \\
  \label{e:wG}
  \wG(\vy,r) &= - \frac{g r}{U} \left[ y_{4} + \ell(\ell+1) y_{3} \right]^{2}.
\end{align}
where $\ell$ is the model harmonic degree, $g$ is the gravity, $N$ the
\BV\ frequency, $S_{\ell}$ the Lamb frequency, and $U$ the usual
homology invariant. The function
\begin{equation}
  \wT(\vy, r) = r^{2} \left[ y_{1}^{2} + \ell(\ell+1) \left( \frac{g}{r \sigma^{2}} \right)^{2} y_{2}^{2} \right]
\end{equation}
appearing in the denominator is proportional to the differential inertia of the mode.

By re-deriving these expressions from equations (13.13) and~(13.15) of
\citet{Unno:1979}\footnote{The equivalent equations in
\citet{Unno:1989} are~(14.16) and~(14.19).}, we find that the correct
expression for $\wG$ is
\begin{equation} \label{e:wG-corr}
   \wG(\vy,r) = - \frac{g r}{U} 
   \left[ y_{4} + (\ell+1) y_{3} \right]^{2};
\end{equation}
comparing this against equation~(\ref{e:wG}), there is an extra factor of
$\ell$ multiplying the $y_{3}$ term in the latter. We have confirmed that this error is typographical in nature; the spurious factor was not included in the calculations presented in \citet{Kawaler:1985} and follow-up papers by the same authors.

\subsection{Numerical Validation}

\begin{figure*}
  \includegraphics{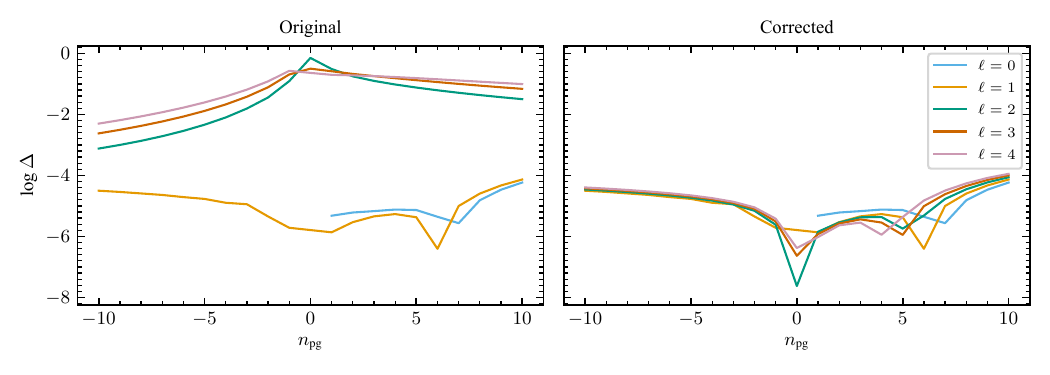}
  \caption{The relative frequency error $\relerr$ between numerical
    and variational frequencies of the $n=3$ polytrope, plotted as a
    function of mode radial order $\npg$ using the original
    (equation~(\ref{e:wG}); left) and corrected
    (equation~(\ref{e:wG-corr}); right) expressions for $\wG$.}
  \label{f:relerr}
\end{figure*}

To validate our analysis we explore how well equation~(\ref{e:var})
can reproduce the numerical eigenfrequencies of an $n=3$ polytrope
model, using both original and corrected expressions for
$\wG$. We apply the \gyre\ oscillation code \citep{Townsend:2013,Townsend:2018,Goldstein:2020}
to calculate adiabatic eigenfrequencies and eigenfunctions for low-radial order,
$\ell=0,\ldots,4$ modes of this model. \gyre's \texttt{VACUUM}
boundary condition (where the Lagrangian pressure perturbation
vanishes at the surface) is adopted, and the oscillation equations are
discretized using the \texttt{COLLOC\_GL2} (second-order
Gauss-Legendre collocation) scheme on a radial grid of $\sim
\num{1000}$ points.

For each mode found we evaluate the relative frequency error
\begin{equation}
  \relerr= \left| \frac{\sigma(\vy)- \sigma}{\sigma} \right|,
\end{equation}
where $\sigma$ is the numerical eigenfrequency reported by
\gyre. Figure~\ref{f:relerr} plots $\relerr$ as a function of mode
radial order $\npg$ \citep[defined in the Eckart-Osaki-Scuflaire
  classification scheme; see][]{Unno:1989}, for calculations based on
the original (equation~\ref{e:wG}) and corrected
(equation~\ref{e:wG-corr}) expressions for $\wG$. In the former case
(left panel) the errors are significant for $\ell \ge 2$ modes,
reaching $\relerr \ga 0.1$ at small $|\npg|$. In the latter
(right panel) the frequency errors are small for all modes, $\relerr
\la \num{e-4}$, and arise primarily from the truncation error of
\gyre's numerical scheme. This confirms the validity of the corrected
expression.

The original expression for $\wG$ happens to yield small $\relerr$ for
both radial ($\ell=0$) and dipole ($\ell=1$) modes. When $\ell=1$ the
spurious factor of $\ell$ in equation~(\ref{e:wG}) makes no difference
to the value of $\wG$. When $\ell=0$, this factor introduces an
additive offset
\begin{equation}
  \delta\wG (\vy) = \frac{gr}{U} y_{3} \left( y_{3} + 2 y_{4} \right)
\end{equation}
to $\wG$; however, integration by parts, together with the fact that
$y_{3}=0$ at the outer boundary for radial modes, can be used to
demonstrate that this offset makes no net contribution to the
numerator of the variational expression~(\ref{e:var}).

\section*{Acknowledgments}

This work has been supported by NSF grants ACI-1663696, AST-1716436
and PHY-1748958, and NASA grant 80NSSC20K0515.

\facilities{We have made extensive use of NASA's Astrophysics Data
  System Bibliographic Services.}  \software{Astropy
  \citep{astropy:2013,astropy:2018,astropy:2022},
  \gyre\ \citep{Townsend:2013,Townsend:2018,Goldstein:2020},
  Matplotlib \citep{Hunter:2007}}


\bibliography{weight-func}

\begin{thebibliography}{}
\expandafter\ifx\csname natexlab\endcsname\relax\def\natexlab#1{#1}\fi
\providecommand{\url}[1]{\href{#1}{#1}}
\providecommand{\dodoi}[1]{doi:~\href{http://doi.org/#1}{\nolinkurl{#1}}}
\providecommand{\doeprint}[1]{\href{http://ascl.net/#1}{\nolinkurl{http://ascl.net/#1}}}
\providecommand{\doarXiv}[1]{\href{https://arxiv.org/abs/#1}{\nolinkurl{https://arxiv.org/abs/#1}}}

\bibitem[{{Astropy Collaboration} {et~al.}(2013){Astropy Collaboration},
  {Robitaille}, {Tollerud}, {Greenfield}, {Droettboom}, {Bray}, {Aldcroft},
  {Davis}, {Ginsburg}, {Price-Whelan}, {Kerzendorf}, {Conley}, {Crighton},
  {Barbary}, {Muna}, {Ferguson}, {Grollier}, {Parikh}, {Nair}, {Unther},
  {Deil}, {Woillez}, {Conseil}, {Kramer}, {Turner}, {Singer}, {Fox}, {Weaver},
  {Zabalza}, {Edwards}, {Azalee Bostroem}, {Burke}, {Casey}, {Crawford},
  {Dencheva}, {Ely}, {Jenness}, {Labrie}, {Lim}, {Pierfederici}, {Pontzen},
  {Ptak}, {Refsdal}, {Servillat}, \& {Streicher}}]{astropy:2013}
{Astropy Collaboration}, {Robitaille}, T.~P., {Tollerud}, E.~J., {et~al.} 2013,
  \aap, 558, A33

\bibitem[{{Astropy Collaboration} {et~al.}(2018){Astropy Collaboration},
  {Price-Whelan}, {Sip{\H{o}}cz}, {G{\"u}nther}, {Lim}, {Crawford}, {Conseil},
  {Shupe}, {Craig}, {Dencheva}, {Ginsburg}, {VanderPlas}, {Bradley},
  {P{\'e}rez-Su{\'a}rez}, {de Val-Borro}, {Aldcroft}, {Cruz}, {Robitaille},
  {Tollerud}, {Ardelean}, {Babej}, {Bach}, {Bachetti}, {Bakanov}, {Bamford},
  {Barentsen}, {Barmby}, {Baumbach}, {Berry}, {Biscani}, {Boquien}, {Bostroem},
  {Bouma}, {Brammer}, {Bray}, {Breytenbach}, {Buddelmeijer}, {Burke},
  {Calderone}, {Cano Rodr{\'\i}guez}, {Cara}, {Cardoso}, {Cheedella}, {Copin},
  {Corrales}, {Crichton}, {D'Avella}, {Deil}, {Depagne}, {Dietrich}, {Donath},
  {Droettboom}, {Earl}, {Erben}, {Fabbro}, {Ferreira}, {Finethy}, {Fox},
  {Garrison}, {Gibbons}, {Goldstein}, {Gommers}, {Greco}, {Greenfield},
  {Groener}, {Grollier}, {Hagen}, {Hirst}, {Homeier}, {Horton}, {Hosseinzadeh},
  {Hu}, {Hunkeler}, {Ivezi{\'c}}, {Jain}, {Jenness}, {Kanarek}, {Kendrew},
  {Kern}, {Kerzendorf}, {Khvalko}, {King}, {Kirkby}, {Kulkarni}, {Kumar},
  {Lee}, {Lenz}, {Littlefair}, {Ma}, {Macleod}, {Mastropietro}, {McCully},
  {Montagnac}, {Morris}, {Mueller}, {Mumford}, {Muna}, {Murphy}, {Nelson},
  {Nguyen}, {Ninan}, {N{\"o}the}, {Ogaz}, {Oh}, {Parejko}, {Parley}, {Pascual},
  {Patil}, {Patil}, {Plunkett}, {Prochaska}, {Rastogi}, {Reddy Janga},
  {Sabater}, {Sakurikar}, {Seifert}, {Sherbert}, {Sherwood-Taylor}, {Shih},
  {Sick}, {Silbiger}, {Singanamalla}, {Singer}, {Sladen}, {Sooley},
  {Sornarajah}, {Streicher}, {Teuben}, {Thomas}, {Tremblay}, {Turner},
  {Terr{\'o}n}, {van Kerkwijk}, {de la Vega}, {Watkins}, {Weaver}, {Whitmore},
  {Woillez}, {Zabalza}, \& {Astropy Contributors}}]{astropy:2018}
{Astropy Collaboration}, {Price-Whelan}, A.~M., {Sip{\H{o}}cz}, B.~M., {et~al.}
  2018, \aj, 156, 123

\bibitem[{{Astropy Collaboration} {et~al.}(2022){Astropy Collaboration},
  {Price-Whelan}, {Lim}, {Earl}, {Starkman}, {Bradley}, {Shupe}, {Patil},
  {Corrales}, {Brasseur}, {N{\"o}the}, {Donath}, {Tollerud}, {Morris},
  {Ginsburg}, {Vaher}, {Weaver}, {Tocknell}, {Jamieson}, {van Kerkwijk},
  {Robitaille}, {Merry}, {Bachetti}, {G{\"u}nther}, {Aldcroft},
  {Alvarado-Montes}, {Archibald}, {B{\'o}di}, {Bapat}, {Barentsen},
  {Baz{\'a}n}, {Biswas}, {Boquien}, {Burke}, {Cara}, {Cara}, {Conroy},
  {Conseil}, {Craig}, {Cross}, {Cruz}, {D'Eugenio}, {Dencheva}, {Devillepoix},
  {Dietrich}, {Eigenbrot}, {Erben}, {Ferreira}, {Foreman-Mackey}, {Fox},
  {Freij}, {Garg}, {Geda}, {Glattly}, {Gondhalekar}, {Gordon}, {Grant},
  {Greenfield}, {Groener}, {Guest}, {Gurovich}, {Handberg}, {Hart},
  {Hatfield-Dodds}, {Homeier}, {Hosseinzadeh}, {Jenness}, {Jones}, {Joseph},
  {Kalmbach}, {Karamehmetoglu}, {Ka{\l}uszy{\'n}ski}, {Kelley}, {Kern},
  {Kerzendorf}, {Koch}, {Kulumani}, {Lee}, {Ly}, {Ma}, {MacBride}, {Maljaars},
  {Muna}, {Murphy}, {Norman}, {O'Steen}, {Oman}, {Pacifici}, {Pascual},
  {Pascual-Granado}, {Patil}, {Perren}, {Pickering}, {Rastogi}, {Roulston},
  {Ryan}, {Rykoff}, {Sabater}, {Sakurikar}, {Salgado}, {Sanghi}, {Saunders},
  {Savchenko}, {Schwardt}, {Seifert-Eckert}, {Shih}, {Jain}, {Shukla}, {Sick},
  {Simpson}, {Singanamalla}, {Singer}, {Singhal}, {Sinha}, {Sip{\H{o}}cz},
  {Spitler}, {Stansby}, {Streicher}, {{\v{S}}umak}, {Swinbank}, {Taranu},
  {Tewary}, {Tremblay}, {Val-Borro}, {Van Kooten}, {Vasovi{\'c}}, {Verma}, {de
  Miranda Cardoso}, {Williams}, {Wilson}, {Winkel}, {Wood-Vasey}, {Xue},
  {Yoachim}, {Zhang}, {Zonca}, \& {Astropy Project
  Contributors}}]{astropy:2022}
{Astropy Collaboration}, {Price-Whelan}, A.~M., {Lim}, P.~L., {et~al.} 2022,
  \apj, 935, 167

\bibitem[{{Brassard} {et~al.}(1992{\natexlab{a}}){Brassard}, {Fontaine},
  {Wesemael}, \& {Hansen}}]{Brassard:1992a}
{Brassard}, P., {Fontaine}, G., {Wesemael}, F., \& {Hansen}, C.~J.
  1992{\natexlab{a}}, \apjs, 80, 369

\bibitem[{{Brassard} {et~al.}(1992{\natexlab{b}}){Brassard}, {Fontaine},
  {Wesemael}, \& {Tassoul}}]{Brassard:1992b}
{Brassard}, P., {Fontaine}, G., {Wesemael}, F., \& {Tassoul}, M.
  1992{\natexlab{b}}, \apjs, 81, 747

\bibitem[{{C{\'o}rsico} \& {Benvenuto}(2002)}]{Corsico:2002}
{C{\'o}rsico}, A.~H., \& {Benvenuto}, O.~G. 2002, \apss, 279, 281

\bibitem[{{Dziembowski}(1971)}]{Dziembowski:1971}
{Dziembowski}, W.~A. 1971, \actaa, 21, 289

\bibitem[{{Goldstein} \& {Townsend}(2020)}]{Goldstein:2020}
{Goldstein}, J., \& {Townsend}, R.~H.~D. 2020, \apj, 899, 116

\bibitem[{Hunter(2007)}]{Hunter:2007}
Hunter, J.~D. 2007, Computing in Science \& Engineering, 9, 90

\bibitem[{{Kawaler} {et~al.}(1985){Kawaler}, {Winget}, \&
  {Hansen}}]{Kawaler:1985}
{Kawaler}, S.~D., {Winget}, D.~E., \& {Hansen}, C.~J. 1985, \apj, 295, 547

\bibitem[{{Kawaler} {et~al.}(1986){Kawaler}, {Winget}, {Hansen}, \&
  {Iben}}]{Kawaler:1986}
{Kawaler}, S.~D., {Winget}, D.~E., {Hansen}, C.~J., \& {Iben}, I., J. 1986,
  \apjl, 306, L41

\bibitem[{{Timmes} {et~al.}(2018){Timmes}, {Townsend}, {Bauer}, {Thoul},
  {Fields}, \& {Wolf}}]{Timmes:2018}
{Timmes}, F.~X., {Townsend}, R. H.~D., {Bauer}, E.~B., {et~al.} 2018, \apjl,
  867, L30

\bibitem[{{Townsend} {et~al.}(2018){Townsend}, {Goldstein}, \&
  {Zweibel}}]{Townsend:2018}
{Townsend}, R.~H.~D., {Goldstein}, J., \& {Zweibel}, E.~G. 2018, \mnras, 475,
  879

\bibitem[{{Townsend} \& {Teitler}(2013)}]{Townsend:2013}
{Townsend}, R.~H.~D., \& {Teitler}, S.~A. 2013, \mnras, 435, 3406

\bibitem[{{Unno} {et~al.}(1989){Unno}, {Osaki}, {Ando}, {Saio}, \&
  {Shibahashi}}]{Unno:1989}
{Unno}, W., {Osaki}, Y., {Ando}, H., {Saio}, H., \& {Shibahashi}, H. 1989,
  {Nonradial oscillations of stars}, 2nd edn. (University of Tokyo Press)

\bibitem[{{Unno} {et~al.}(1979){Unno}, {Osaki}, {Ando}, \&
  {Shibahashi}}]{Unno:1979}
{Unno}, W., {Osaki}, Y., {Ando}, H., \& {Shibahashi}, H. 1979, {Nonradial
  oscillations of stars}, 1st edn. (University of Tokyo Press)

\end{thebibliography}

\end{document}